\documentclass[aps,preprint,nofootinbib,showpacs]{revtex4}
\usepackage{mathrsfs}
\usepackage{graphicx}
\usepackage{epsfig}

\def\beq{\begin{eqnarray}}
\def\eeq{\end{eqnarray}}
\def\non{\nonumber}
\def\la{\langle}
\def\ra{\rangle}
\def\Un{{\cal U}}
\def\Re{{\rm Re}}
\def\Im{{\rm Im}}

\begin{document}

\title{Constraints of unparticle physics parameters from $K^0-\bar K^0$ mixing}
\author{Li-Gong Bian\footnote{Email: lgb@mail.nankai.edu.cn},
        Zheng-Tao Wei\footnote{Email: weizt@nankai.edu.cn} }
\affiliation{ School of Physics, Nankai University, Tianjin 300071, China }

\begin{abstract}
The neutral kaon meson mixing plays an important role in test of the Standard Model (SM) and new physics beyond it. Scale invariant unparticle physics induces a flavor changing neutral current (FCNC) transition of $K^0-\bar K^0$ oscillation at the tree level. In this study, we investigate the scale invariant unparticle physics effects on the $K^0-\bar K^0$ mixing. Based on the current experimental data, we give constraints of $K^0-\bar K^0$ mixing on the unparticle parameters.
\end{abstract}



\maketitle

\section{Introduction}

The neutral kaon meson system had played an important role in history. The mixing-induced and indirect CP violation were both first discovered in this system \cite{ccft}. In the SM, the neutral kaon mixing is occurred through a flavor changing neutral current transition depicted by a box diagram at the loop level. Thus it provides an important place to test the SM and/or explore new physics beyond it. The interests in research of new physics in the neutral kaon system is very long. Even recently, there are many new physics studies, for example,  in \cite{Buras:2008nn,Laiho:2009eu,Gedalia,Trine,Zhang,Duling}. At present, the experimental data on the neutral kaon mixing is very precise, and the theoretical studies, in particular the lattice calculations on the non-perturbative quantities are also improved a lot. The researches on the neutral kaon system are going into a deeper level.

The purpose of the study is to explore the neutral kaon mixing within a new physics scenario called by unparticle. This is an idea proposed by Georgi in \cite{Georgiek,Georgi2007si}. Scale invariance is the guiding principle in this scenario.  A scale invariant stuff, named by unparticle, possesses some properties which are different from those of the ordinary particles. The dimension of unparticle is in general fractional rather than an integral number. Another aspect is that the real unparticle has no definite mass. The interactions between the unparticle and the SM particles are described in the framework of low energy effective theory and lead to various interesting phenomena. The unparticle physics had been got intensive interests after the idea was proposed. Although this topic is currently not hot, it is still necessary to study its effects in different physical processes. Because we don't know what is the right direction of new physics and which new physics model favors the real world.

Within the unparticle scenario, the neutral meson mixing, such as $D^0-\bar D^0$, $B_d^0-\bar B^0_d$ and $B_s^0-\bar B_s^0$ have been explored \cite{Luo,Li,Chen,ChenH}. Only the neutral kaon mixing is not studied. Because the energy scale related to the kaon system is lower than other heavy mesons, the SM contribution should be more dominant and the new physics will play less important role. However, the more precise data on the kaon system can provide more stringent constraint on the new physics parameters. This is the reason why the kaon system is still an active research area. One difficult problem related to the kaon system is the long distance contribution to the mixing parameter which is not easy to evaluate. This problem becomes more serious in the kaon system than that in the heavy meson. The aim of this study is to constrain the unparticle parameters from the neutral kaon mixing. Because there is no similar study before, it is necessary to investigate whether the unparticle scenario can be applicable to the kaon system.

\section{$K^0-\bar K^0$ mixing in the SM}

At first, we give notations for the neutral kaon system. In the $K^0-\bar K^0$ system, the oscillation between the two neutral kaon mesons is described by the equation
 \beq
  i\frac{d\psi(t)}{dt}=\hat H \psi(t),
  \qquad \psi(t)=\left(\begin{array}{c} |K^0(t)\rangle \\
  |\bar K^0(t)\rangle \end{array}\right),
 \eeq
where
 \beq
  \hat H=\hat M-i\frac{\hat\Gamma}{2} = \left(\begin{array}{cc}
   M-i\frac{\Gamma}{2} & M_{12}-i\frac{\Gamma_{12}}{2} \\
   M_{12}^*-i\frac{\Gamma_{12}^*}{2}  & M-i\frac{\Gamma}{2}
   \end{array}\right).
 \eeq
with $M_{ij}$ and $\Gamma_{ij}$ being the transition matrix elements. We have assumed the CPT conservation and hermiticity for the matrices $\hat M$ and $\hat\Gamma$.

After diagonalizing the system and using the convention $CP|K^0\rangle=-|\bar K^0\rangle$, $CP|\bar K^0\rangle=-|K^0\rangle$, we obtain the eigenstates:
 \beq \label{KLS}
 |K_{L,S}\rangle=\frac{(1+\bar\varepsilon)|K^0\rangle\pm (1-\bar\varepsilon)|\bar K^0\rangle}
  {\sqrt{2\left(1+\mid\bar\varepsilon\mid^2\right)}},
 \eeq
where the mixing parameter $\bar\varepsilon$ is a small complex quantity given by
 \beq
 \frac{1-\bar\varepsilon}{1+\bar\varepsilon}=
  \sqrt{\frac{M^*_{12}-i\frac{1}{2}\Gamma^*_{12}}
  {M_{12}-i\frac{1}{2}\Gamma_{12}}}.
 \eeq
In the SM, the oscillation between the flavor states $K^0$ and $\bar K^0$ are caused by weak interactions, thus the above eigenstates are called by the weak eigenstates. In new physics which is beyond the SM, the eigenstates represent generalized states including both the SM and new physics effects.

The eigenvalues associated with the eigenstates are
 \begin{equation}
  M_{L,S}=M\pm \Re Q,   \qquad
  \Gamma_{L,S}=\Gamma\mp 2 \Im Q,
 \end{equation}
where
 \beq
  Q=\sqrt{\left(M_{12}-i\frac{1}{2}\Gamma_{12}\right)
   \left(M^*_{12}-i\frac{1}{2}\Gamma^*_{12}\right)}.
 \eeq
We thus get
 \beq
 \Delta M &=&M(K_{L})-M(K_{S})= 2{\rm Re} Q, \non\\
 \Delta\Gamma&=&\Gamma(K_{L})-\Gamma(K_{S})=-4 {\rm Im} Q.
 \eeq

Since $\bar\varepsilon$ is small, at the order of ${\cal O}(10^{-3})$, the below relations are reasonable,
 \beq\label{eq:ReIm}
 {\rm Im} M_{12} \ll {\rm Re} M_{12}, \qquad
 {\rm Im} \Gamma_{12} \ll {\rm Re} \Gamma_{12}.
 \eeq
It should be noted that the above relations are still applicable in the unparticle physics. Thus, neglecting the small imaginary part of $M_{12}$ and $\Gamma_{12}$, we can get a good approximation:
 \beq\label{eq:mass_difference}
 \Delta M_{K} \cong 2 {\rm Re} M_{12}, \qquad
 \Delta\Gamma_{K}\cong 2 {\rm Re}\Gamma_{12}.
 \eeq
In the SM, the matrix elements $M_{12},\Gamma_{12}$ of the $K^0-\bar K^0$ mixing contains both the short distance (SD) and long distance (LD) contributions:
\beq
  M_{12}^{SM}= M_{12}^{SD}+ M_{12}^{LD}, \qquad
 \Gamma_{12}^{SM}=\Gamma_{12}^{SD}+ \Gamma_{12}^{LD}.
 \eeq
An important formula for $\bar\varepsilon$ including both the SD and LD contributions of the SM is \cite{Buras:2010pza}
 \beq \label{eq:ep}
 \bar\varepsilon=\kappa_{\varepsilon}\frac{e^{i\phi_{\varepsilon}}}{\sqrt{2}}\frac{{\rm Im} M_{12}}
 {\Delta M_{K}}.
\eeq
Where the phase $\phi_{\varepsilon}=(43.51\pm0.05)^{\circ}$ and factor $\kappa_{\varepsilon}=0.94\pm0.02$.  Thus, the mixing parameter $\bar\varepsilon$ is approximated by
 \beq \label{eq:ep}
 \bar\varepsilon\approx e^{i\phi_{\varepsilon}}\sin\phi_{\varepsilon}\frac{{\rm Im} M_{12}}{\Delta M_{K}}.
 \eeq

At the quark level, the flavor changing neutral current transitions between $K^0$ and $\bar K^0$ are induced through a box diagram with exchange of the intermediate up type quarks. This is the the short distance  contribution to $M_{12}$. From \cite{Burdman}, the formula is given as
 \beq \label{eq:SM}
 M_{12}^{\rm SD} = \frac{G_{\rm F}^{2}}{12 \pi^{2}}
  f_{K}^{2}\hat{B}_{K}m_{K}m_{W}^{2}\left[\lambda_{c}^{*2}
  \eta_{1}S_{0}(x_{c})+\lambda_{t}^{*2}\eta_{2}S_{0}
  (x_{t})+2\lambda_{c}^{*}\lambda_{t}^{*}\eta_{3}S_{0}(x_{c},x_{t})\right],
 \eeq
where $f_{K}$ is the $K$-meson decay constant, $m_{K}$ is the $K$-meson
mass, and $\hat B_K$ is the renormalization group invariant parameter. The
parameters $\lambda_{i} = V_{is}^{*} V_{id}$ where $V_{is}$ and $
V_{id}$ are the CKM matrix elements. The functions $S_0$ are
 \beq
 S_0(x_{t})&=&\frac{4x_{t}-11x_{t}^{2}+x_{t}^{3}}{4(1-x_{t})^{2}}-
  \frac{3x_{t}^{3} \ln x_{t}}{2(1-x_{t})^{3}}, \non\\
 S_0(x_{c})&=&x_{c}, \\
 S_0(x_{c},x_{t})&=&x_{c}\left[\ln\frac{x_{t}}{x_{c}}-\frac{3x_{t}}{4(1-x_{t})}-
 \frac{3 x^{2}_{t}\ln x_{t}}{4(1-x_{t})^{2}}\right], \non
 \eeq
The values of $\eta_{i}$ are taken to be the next-to-leading-order (NLO) results
given in \cite{Buras,Herrlichhh,Herrlichvf,Herrlichyv}
 \beq
 \eta_{1}=1.38\pm0.20,~
 \eta_{2}=0.57\pm0.01,~
 \eta_{3}=0.47\pm0.04.
 \eeq

For the $\Gamma_{12}$, the SD contribution arises from the virtual quark interactions which is expected to be very small. The LD contribution coming from the intermediate hadron states should dominate. However, the theoretical uncertainties due to the non-perturbative dynamics is very large. Usually, it's difficult to separate the hadron uncertainties and the new physics effects. As will be shown, the unparticle parameters can be constrained from the term $M_{12}$. So, in the discussion later,  we will not use the $\Gamma_{12}$ in order to reduce the theoretical uncertainties.

In the above derivations and the forthcoming discussions in the
unparticle physics, the below relations are useful
 \beq
 & &\la\bar{K^0}|\bar s\gamma_\mu(1-\gamma_5)d\bar s\gamma^\mu(1-\gamma_5)d
  |K^0\ra=\frac{8}{3}f_K^2m_K^2\hat B_K, \non \\
& &\la\bar{K^0}|\bar s(1-\gamma_5)d\bar s(1-\gamma_5)d|K^0\ra =
  -\frac{5}{3}f_K^2m_K^2\hat B_K
  (\frac{m_{K}}{m_{s}+m_{d}})^{2}.
 \eeq
where $m_{s,d}$ are the strange and down quark masses.

\section{$K^0-\bar K^0$ mixing in unparticle physics}

In this section, we turn to study the $K^0 -\bar K^0$ mixing in unparticle physics. In the low energy effective theory, unparticle fields will emerge below an energy scale $\Lambda_\Un$ \cite{Georgiek}. The relevant low energy effective interaction for the $s$ and $d$ quarks is described by
 \beq
 \frac{C_S^{ds}}{\Lambda_\Un^{d_\Un}}
  \bar s\gamma_{\mu}(1-\gamma_5)d~\partial^\mu O_\Un+
 \frac{C_V^{ds}}{\Lambda_\Un^{d_\Un-1}}
  \bar s\gamma_{\mu}(1-\gamma_5)d~O_\Un^\mu+h.c.
 \eeq
where $O_\Un$ and $O_\Un^{\mu}$ denote operators of the scalar and vector unparticle fields respectively. The $C_S$ and $C_V$ are dimensionless coefficients and they depend on quark and lepton flavors in general. Since only quarks $s$ and $d$ are concerned in this study, we will simplify $C_S^{ds}\to C_S$ and $C_V^{ds}\to C_V$ in the later discussions.

The propagators for the scalar and vector unparticle fields with the time-like momentum $P$ are given by \cite{Georgi2007si,Grinstein2008qk}
 \beq
 &&\int d^4 x e^{iP\cdot x}\la 0 |TO_\Un(x)O_\Un(0)|0\ra \non\\
 &&=
  i\frac{A_{d_\Un}}{2{\rm sin}(d_\Un\pi)}\frac{1}{(P^2+i\epsilon)^{2-d_\Un}}
   e^{-id_\Un\pi}, \non\\
 &&\int d^4 x e^{iP\cdot x}\la 0 |TO^{\mu}_\Un(x)O^{\nu}_\Un(0)|0\ra \non\\
 &&=
  i\frac{A_{d_\Un}}{2{\rm sin}(d_\Un\pi)}\frac{-g^{\mu\nu}+2(d_\Un-2)P^{\mu}P^{\nu}/(d_\Un-1)P^{2}}
  {(P^{2}+i\epsilon)^{2-d_\Un}}e^{-id_\Un\pi},
 \eeq
where
 \beq
 A_{d_\Un}=\frac{16\pi^{5/2}}{(2\pi)^{2d_\Un}}
  \frac{\Gamma(d_\Un+1/2)}{\Gamma(d_\Un-1)\Gamma(2d_\Un)}.
 \eeq
Here we consider the vector unparticle which is not transverse: $\partial_\mu O_\Un^\mu\neq0$ unless $d_\Un=0$ due to unitarity constraint \cite{Mack:1975je}. Another property of $d_\Un$ requested by the unitarity constrain is that $d_\Un\geq 3$ for vector unparticles and $d_\Un\geq 1$ for scalar unparticles \cite{Grinstein2008qk}. The scale dimension $d_\Un$ is fractional in general, and it cannot be integral (except $d_\Un=1$. where the singularity is canceled by $A_{d_\Un}$ in the nominator) due to the singularity of the function ${\rm sin}(d_\Un\pi)$ in the
denominator. The factor $e^{-id_\Un\pi}$ provides a CP conserving phase which produces peculiar interference effects in high energy scattering processes \cite{Georgi2007si}, Drell-Yan process
\cite{Cheung} and CP violation in B decays \cite{Chen}.

\begin{figure}[!htb]
\begin{center}
\begin{tabular}{cc}
\includegraphics[width=5in]{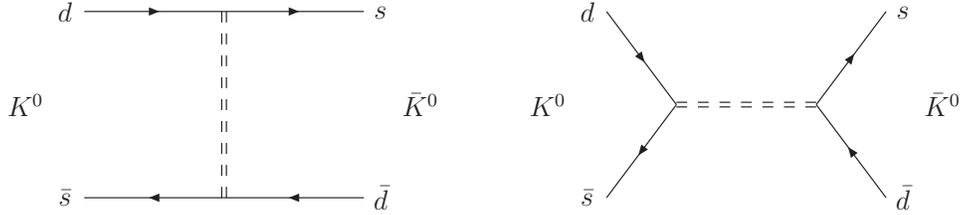}
\end{tabular}
\end{center}
\caption{The Feynman graphs of the $K^{0}-\bar{K^{0}}$ mixing in
 unparticle physics. The double dashed lines represent the unparticle
 fields. } \label{fig1}
\end{figure}

Unlike the SM where the $\Delta S=2$ FCNC transitions occurred at loop level, the unparticle contribution can induce a tree level FCNC transitions between $K^0$ and $\bar K^0$ which is depicted in Fig. \ref{fig1}. There are two diagrams corresponding to t- and s-channel unparticle exchanges. For the s-channel, the momentum of the unparticle $P^2=m_K^2$. For the t-channel, the the momentum of the unparticle is not fixed which introduces an uncertainty in the calculation. Because strange quark mass is much greater than that of the down quark, $m_s\gg m_d$, the momentum of the kaon meson is mostly carried by the strange quark. The momentum of the exchanged unparticle will be at the order of $m_K$, and we make an approximation that $P^2\approx m_K^2$.

Using the Feynman rules given above, we obtain the final expression for the transition matrices in the unparticle physics. For the scalar unparticle,
 \beq \label{eq:S1}
 &&M_{12}^{\Un}=\frac{5C_{S}^{2}}{12}\frac{f_{K}^{2}\hat B_K
  A_{d_\Un}}{m_{K}}\left(\frac{m_{K}}{\Lambda_{\Un}}\right)^{2d_\Un}
  \left(\frac{m_{K}}{m_{s}+m_{d}}\right)^{2}{\rm cot}(d_\Un\pi), \non\\
 &&\Gamma_{12}^{\Un}=\frac{5C_{S}^{2}}{6}\frac{f_{K}^{2}\hat B_K
  A_{d_\Un}}{m_{K}}\left(\frac{m_{K}}{\Lambda_{\Un}}\right)^{2d_\Un}
  \left(\frac{m_{K}}{m_{s}+m_{d}}\right)^{2},
 \eeq
and
 \beq \label{eq:V1}
 M_{12}^{\Un}&=&\frac{C_{V}^{2}}{4}\frac{f_{K}^{2}\hat B_K
  A_{d_\Un}}{m_{K}}\bigg(\frac{m_{K}}{\Lambda_{u}}\bigg)^{2d_\Un-2}\bigg[\frac{8}
  {3}-\frac{10(d_\Un-2)}{3(d_\Un-1)}\bigg(\frac{m_{K}}{m_{s}+m_{d}}\bigg)^{2}
  \bigg]{\rm cot}(d_\Un\pi), \non\\
 \Gamma_{12}^{\Un}&=&\frac{C_{V}^{2}}{2}\frac{f_{K}^{2}\hat B_K
  A_{d_\Un}}{m_{K}}\bigg(\frac{m_{K}}{\Lambda_{u}}\bigg)^{2d_\Un-2}
  \bigg[\frac{8}{3}-\frac{10(d_\Un-2)}{3(d_\Un-1)}\bigg(\frac{m_{K}}{m_{s}+m_{d}}\bigg)^{2}
  \bigg].
 \eeq
for the vector unparticle. For both cases, we have
 \beq
 M_{12}^{\Un}=\frac{\Gamma_{12}^{\Un}}{2}~{\rm cot}(d_\Un\pi).
 \eeq
The above relation has been given in \cite{ChenH}.

For the mixing parameter $\bar\varepsilon$, it is straightforwardly
to obtain
 \beq \label{eq:ep2}
 \bar\varepsilon^{\Un}=e^{i\phi_{\varepsilon}}\sin\phi_{\varepsilon}\frac{{\rm Im} M_{12}^{\Un}}{\Delta  M_{K}^{\rm exp}}.
 \eeq
where we have used the $\Delta  M_{K}^{\rm exp}$ to replace the $\Delta M_{K}$ in Eq.~(\ref{eq:ep}). The above formula is applicable for both the scalar and vector unparticles.

\section{Constraints of the unparticle parameters }

In the $K^{0}-\bar{K^{0}}$ system, the mass difference $\Delta M_{K}$, width difference $\Delta \Gamma_{K}$ and the CP violating parameter $\bar{\varepsilon}$ are the most important experimental parameters. We will make use of these data to constrain the phenomenological parameters of the unparticle physics. From PDG \cite{Nakamura}, the experimental results are
 \beq
 \Delta M_{K}^{\rm exp}&=&( 3.483\pm0.006)\times10^{-15}~{\rm GeV},\non\\
 |\bar\varepsilon|^{\rm exp}&=&( 2.228\pm0.001 )\times 10^{-3}.
 \eeq
The relations between the experiment and theory have been obtained as
 \beq
 &&\Delta M_K^{\rm exp}=2{\rm Re}\left(M_{12}^{\rm SD}
  +M_{12}^{\rm LD}+M_{12}^{\Un}\right),\non\\
 &&|\bar\varepsilon|^{\rm exp}=\frac{|e^{i\phi_{\varepsilon}}|}{\Delta  M_{K}^{\rm exp}}
 \left[\frac{\kappa_{\varepsilon}}{\sqrt{2}}{\rm Im}M_{12}^{SD}+\sin\phi_{\varepsilon}{\rm Im}M_{12}^{\Un}
  \right].
 \eeq

The SD contribution of $M_{12}^{\rm SM}$ is given in Eq. (\ref{eq:SM}) and can be calculated reliably. The main uncertainty in the SM comes from the parameter $\hat B_K$. We use the value $\hat B_K=0.724\pm 0.024$ given in \cite{Colangelo:2010et} calculated from a lattice method.
For the LD part ${\rm Re}M_{12}^{LD}$, it is taken from \cite{Buras:2010pza}:
 \beq
 {\rm Re}M_{12}^{LD}=\frac{\Delta M_K^{\rm LD}}{2}, \qquad
 \frac{\Delta M_K^{\rm LD}}{\Delta M_K^{\rm exp}}\approx0.1\pm0.2.
 \eeq

After subtracting the SM contribution, the remained is the new physics effect. Thus, we can know the unparticle contributions ${\rm Re}M_{12}^{\Un}$ and ${\rm Im}M_{12}^{\Un}$. Using Eqs. (\ref{eq:S1}, \ref{eq:V1}), we obtain the constraints of $C_{S}$, $C_{V}$ as following. For the coupling coefficients $C_{S}^{2}$ and $C_{V}^{2}$, they are obtained as
\beq \label{eq:icsv3}
C_{S}^{2}&=&\frac{12(m_{s}+m_{d})^2~{\rm sin}(d_\Un\pi)}{5A_{d_\Un}\hat B_K f_K^{2}m_K}
  \left(\frac{m_{K}}{\Lambda_{\Un}}\right)^{-2d_{\Un}}~M_{12}^{\Un}, \non\\
 C_{V}^{2}&=&\frac{4m_K~{\rm sin}(d_\Un\pi)}{A_{d_\Un}\hat B_K f_{K}^{2}\left[\frac{8}{3}-\frac{10(d_\Un-2)}{3(d_\Un-1)}
  (\frac{m_{K}}{m_{s}+m_{d}})^{2}\right]}\left(\frac{m_{K}}{\Lambda_{\Un}}\right)^{2-2d_\Un}
  M_{12}^{\Un}.
 \eeq
one need to note that Eqs.~(\ref{eq:ReIm},\ref{eq:mass_difference}) are used in deriving above equations. 
The vector coupling is more suppressed due to an additional factor $\left(\frac{m_{K}}{\Lambda_{\Un}}\right)^2$. Similarly, the real and imaginary parts of the scalar coupling $C_{S}^{2}$ are given by
 \beq \label{eq:rcsv1}
 {\rm Re}C_{S}^{2}&=&\frac{12(m_{s}+m_{d})^2~{\rm tan}(d_\Un\pi)}{5A_{d_\Un}\hat B_K
 f_K^{2}m_K}\left(\frac{m_{K}}{\Lambda_{\Un}}\right)^{-2d_{\Un}}~
  {\rm Re}M_{12}^{\Un}, \non\\
  {\rm Im}C_{S}^{2}&=&\frac{12(m_{s}+m_{d})^2~{\rm tan}(d_\Un\pi)}{5A_{d_\Un}\hat B_K
 f_K^{2}m_K}\left(\frac{m_{K}}{\Lambda_{\Un}}\right)^{-2d_{\Un}}~
  {\rm Im}M_{12}^{\Un},
 \eeq
 and for the vector coupling $C_V^{2}$,  the real and imaginary parts are given by
 \beq \label{eq:icsv2}
  {\rm Re}C_{V}^{2}&=&\frac{4m_K{\rm tan}(d_\Un\pi)}{A_{d_\Un}\hat B_K
 f_{K}^{2}\left[\frac{8}{3}-\frac{10(d_\Un-2)}{3(d_\Un-1)}(\frac{m_{K}}
 {m_{s}+m_{d}})^{2}\right]}\left(\frac{m_{K}}{\Lambda_{\Un}}\right)^{2-2d_\Un}~
  {\rm Re}M_{12}^{\Un},\non\\
  {\rm Im}C_{V}^{2}&=&\frac{4m_K{\rm tan}(d_\Un\pi)}{A_{d_\Un}\hat B_K f_{K}^{2}\left[\frac{8}{3}-\frac{10(d_\Un-2)}{3(d_\Un-1)}
  (\frac{m_{K}}{m_{s}+m_{d}})^{2}\right]}\left(\frac{m_{K}}{\Lambda_{\Un}}\right)^{2-2d_\Un}~{\rm Im}M_{12}^{\Un}.
 \eeq

\subsection{Input parameters}

Here, we collect all the input parameters used in the numerical analysis. Most parameters are taken from PDG \cite{Nakamura}.

CKM parameters ($A$, $\lambda$, $\overline{\rho}$,
$\overline{\eta}$):
 \beq
 A=0.804_{-0.015}^{+0.022}, && \lambda=0.2253\pm0.0007,  \\
 \overline{\rho}=0.132_{-0.014}^{+0.022}, && \overline{\eta}=0.341\pm0.013.
 \eeq

Decay constant of kaon meson:
 \beq
 f_K=160~{\rm MeV}.
 \eeq

Quark and gauge boson masses:
 \beq
 &&m_d=4.1-5.8~{\rm MeV},  \qquad m_{s}=101_{-21}^{+29}~{\rm MeV},  \nonumber\\
 &&m_c=1.27_{-0.09}^{+0.07}~{\rm GeV}, \qquad m_{W}=80.384\pm0.014~GeV.\nonumber\\
 &&m_{t}=171.2\pm0.9\pm1.3~{\rm GeV}.
 \eeq

\subsection{Bounds on the unparticle parameters}

After the above preparations, we are now ready to discuss the numerical results. The phenomenological parameters of unparticle physics are: scale dimension $d_\Un$, energy scale $\Lambda_\Un$ and the coupling coefficients $C_S(C_V)$. At first, we provide an estimate on the magnitude of the scalar coupling parameter $C_S$ from the neutral kaon mixing and then compare it with the values constrained from the neutral B and D systems. As \cite{Li},  we fix the energy scale and dimension by $\Lambda_\Un=1$~TeV and $d_\Un=3/2$. The numerical results of the absolute value of the scalar coupling parameter $|C_{S}|$ are given in Table \ref{ta1}. We also include the upper bounds of the coupling from neutral B and D systems given in \cite{Li} for comparison. Because the unitarity constraint for the vector unparticle is not considered in \cite{Li}, it is impossible to compare the results for the vector coupling coeffients. Thus, only the scalar coupling parameter is given. From Table \ref{ta1}, the scalar coupling coefficient $C_S$ is constrained to be $C_S=7.1\times 10^{-3}$.  Under the assumption that the coupling parameter is flavor independent, the kaon mixing provides more stringent constraints on unparticle coupling than other systems.

\begin{table}[!ht]
\caption{The constraints on the coupling parameter $|C_{S}|$ with $d_\Un=3/2$ and $\Lambda_\Un=1$ TeV.} \label{ta1}
\begin{ruledtabular}
\begin{tabular}{cccc}
          &  From B-system       & From D-system      & From K-system       \\ \hline
 $|C_S|$ & $3.4\times10^{-2}$  & $2.1\times10^{-2}$ & $7.1\times10^{-3}$
\end{tabular}
\end{ruledtabular}
\end{table}

\begin{figure}
\centerline{\psfig{file=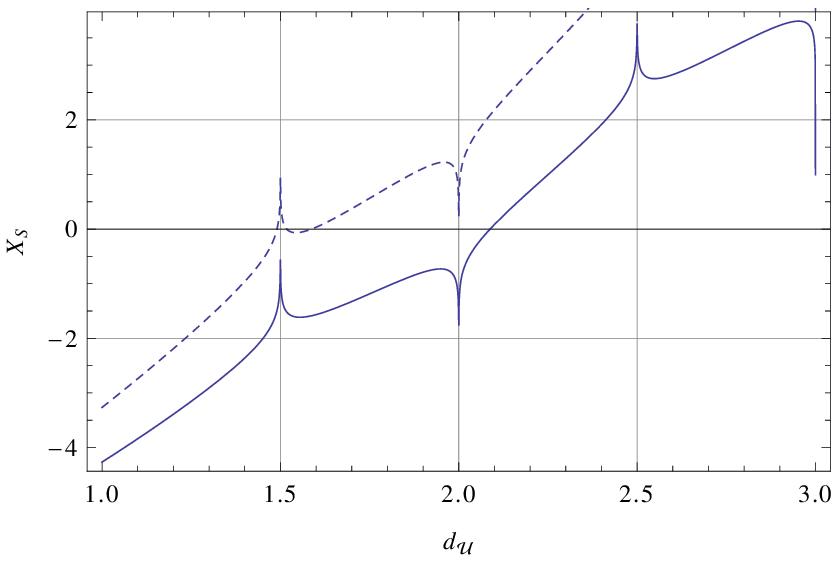,width=10cm}}
\vspace{1.5cm}
\centerline{\psfig{file=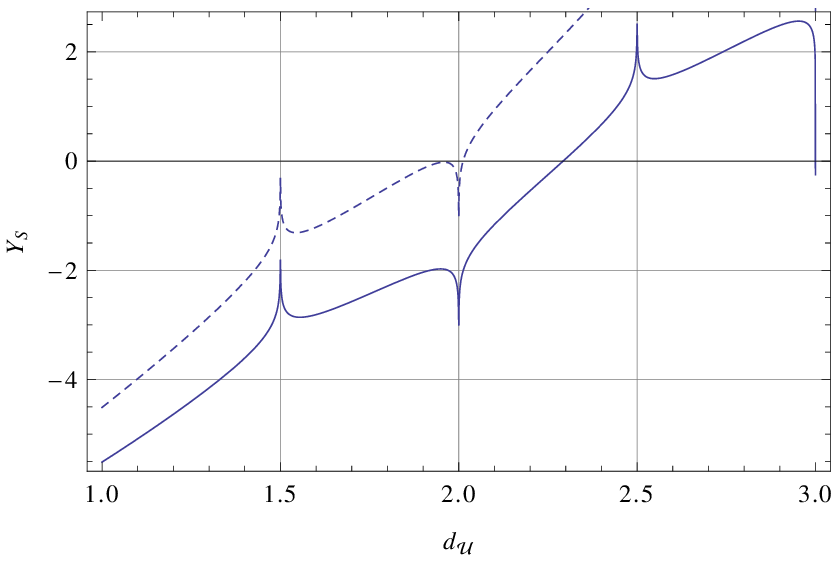,width=10cm}}
\vspace{0.5cm}
\caption{The scalar unparticle coupling parameter $C_{S}$ versus the scale dimension $d_\Un$  where the vertical variable $X_S={\rm log}_{10}\sqrt{|\Re C_{S}^{2}|}$, $Y_S={\rm log}_{10}\sqrt{|\Im C_{S}^{2}|}$. The solid line is given for $\Lambda_{\Un}=1$ TeV and the dashed for $\Lambda_{\Un}=10$~TeV.} \label{fig2}
\end{figure}

\begin{figure}
\centerline{\psfig{file=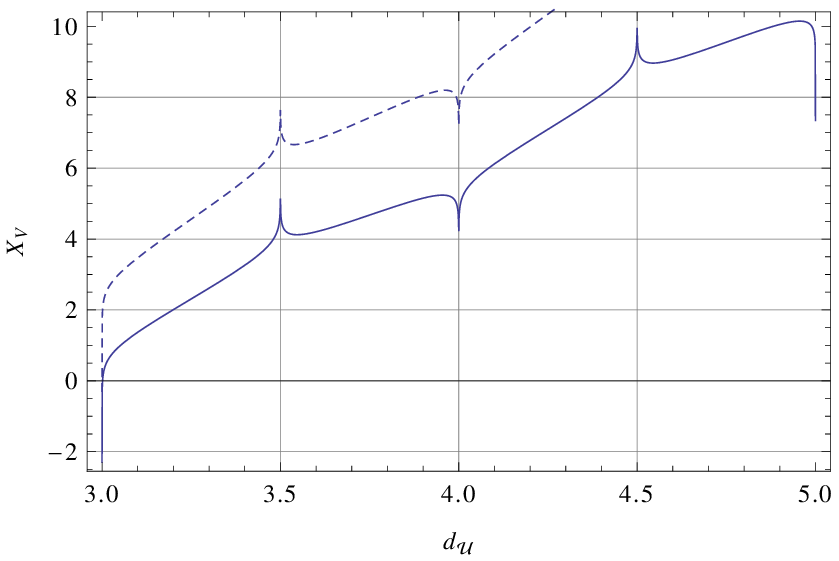,width=10cm}}
\vspace{1.5cm}
\centerline{\psfig{file=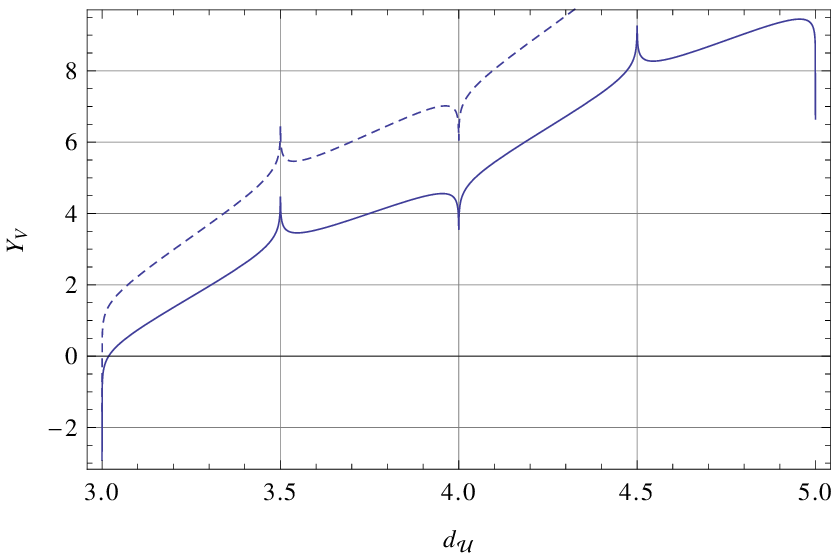,width=10cm}}
\vspace{0.5cm}
\caption{The scalar unparticle coupling parameter $C_{V}$ versus the scale dimension $d_\Un$ where the vertical variable $X_V={\rm log}_{10}\sqrt{|\Re C_{V}^{2}|}$, $Y_V={\rm log}_{10}\sqrt{|\Im C_{V}^{2}|}$.  The solid line is given for $\Lambda_{\Un}=1~TeV$ and the dashed for $\Lambda_{\Un}=10~TeV$}. \label{fig3}
\end{figure}

As a next step, we consider the general case where the scale dimension is varied. At present, the exact knowledge about the range of the scale dimension is still unknown. As we have pointed out before, unitarity constraints request $d_\Un\geq 1$ for scalar and $d_\Un\geq 3$ for vector unparticles. For the coupling parameters $C_S(C_V)$, their real and imaginary parts will be explored separately unlike the analysis in \cite{Li,Chen}. Because the coupling coefficients $C_S(C_V)$ are very sensitive to the scale dimension  $d_\Un$ and change rapidly as $d_\Un$ varies, we introduce logarithmic functions $X_{(S,V)}={\rm log}_{10}\sqrt{|\Re C_{(S,V)}^{2}|}$ and $Y_{(S,V)}={\rm log}_{10}\sqrt{|\Im C_{(S,V)}^{2}|}$ (here the subscripts "V, S" represent the scalar and vector cases).  The numerical results of the the functions $X_{(S,V)}$ and $Y_{(S,V)}$ for the scalar and vector unparticle versus the scale dimension $d_\Un$ are plotted in Figs. \ref{fig2} and \ref{fig3}. The range of the dimension is chosen to be $1<d_\Un<3$ for the scalar and $3<d_\Un<5$ for the vector cases.

Since $X_{(S,V)}$ and $Y_{(S,V)}$ represent the order of $\Re C_{(S,V)}$ and $\Im C_{(S,V)}$, negative values mean that the coupling coefficients are smaller than 1. If the coupling parameters are too large, one may meet the non-perturbative problem. If we require that the coupling parameters are smaller than 1, we obtain $1<d_\Un<2$ for scalar unparticle. For the vector unparticle, $d_\Un$ has to lie very close to 3. This unnatural thing indicates that the vector unparticle contribution is either too small or a large coupling parameter is required. From Figs. \ref{fig2} and \ref{fig3}, it is shown that the values of $X$ are larger than $Y$ by about 1-2, thus the magnitude of the real part of the coupling parameters $C_{(S,V)}$ is larger than their imaginary part by 1-2 orders for both the scalar and vector unparticles. The physical reason is that the real part is proportional to $\Delta M_K^{\rm exp}$ while the imaginary part contributes to the small kaon mixing parameter $\bar\varepsilon$. Another property of these figures is that the parameters  $X$ and $Y$ are increasing in nearly the whole range except at the integral and half integral points of $d_\Un$.  This is because of the $\tan(d_\Un \pi)$ function, will break to $+\infty$ or $-\infty$ when the dimension is getting half integral,  or become to be zero when dimension is integral. In both cases, there will be no constraints on $C_{(S,V)}$. The dependence of $C_{(S,V)}$ on the energy scale $\Lambda_\Un$ is simple because $C_{(S,V)}$ are proportional to $\Lambda_\Un^{d_\Un}$ or $\Lambda_\Un^{d_\Un-1}$. This can be also seen clearly from figures where we gives two cases of $\Lambda_\Un=1$ TeV and $\Lambda_\Un=10$ TeV.

\section{Conclusions and discussions}\label{sec5}

In this study, the new physics effects from scale invariant unparticle sector on the $K^0-\bar K^0$ mixing are explored. The SM contribution, in particular the long distance part, will produce large uncertainties which is not under well control. This difficulty exists for any new physics search in the kaon system. We have considered the long distance contributions in a simple and maybe a bit crude way. This treatment can be improved in the future if we have better knowledge of the non-perturbative hadron dynamics.

With the unparticle scenario, the flavor changing neutral current transitions of $K^0-\bar K^0$ mixing can occur at tree level. Thus, the neutral kaon system provides a sensible probe to unparticle physics. We observe that the kaon system gives very stringent constraint on the parameters. The coupling parameter for the scalar unparticle and quarks is obtained to be at the order of $10^{-3}$. For the vector unparticle, if the unitarity condition is imposed, there is nearly no parameter space to let the coupling coefficient be smaller than 1. When the scale dimension $d_\Un$ is larger than 3, the vector unparticle contribution will be small,  otherwise, large coupling parameters are required which may induce non-perturbative problem. Our numerical results show that the coupling parameters are very sensitive to the choice of the scale dimension.

The unparticle physics effects on the neutral B and D mixing had been studied a lot in literature but no study on the neutral kaon system up to now. Our result show that the neutral kaon is also important for testing the unparticle scenario.

\section*{Acknowledgments}

This work was supported in part by NNSFC under contract No. 11175091.

\end{document}